\documentclass[a4paper,10pt,onecolumn]{npr}
\usepackage{multicol}
\usepackage{graphicx}
\usepackage{amssymb,bm,mathrsfs,bbm,amscd}
\usepackage[tbtags]{amsmath}
\usepackage{lastpage}

\begin{document}

%%  No Modify  %%
%___________________________________________________________________________________
\def\rd{{\rm d}}
\newcommand{\lcf}{f}
\newcommand{\sbar}{\overline{s}}
\newcommand{\tbar}{\overline{t}}
\newcommand{\lam}{\lambda}
\newcommand{\Ht}{\lambda_{\overline{2}}\lambda_{4};\lambda_{1}\lambda_{\overline{3}}}
\newcommand{\Htp}{\lambda'_{\overline{2}}\lambda'_{4};\lambda'_{1}\lambda'_{\overline{3}}}
\newcommand{\Htm}{\lambda'_{\overline{2}}\lambda'_{4}\lambda'_{1}\lambda'_{\overline{3}}}
\newcommand{\Hsm}{\lambda_{3}\lambda_{4}\lambda_{1}\lambda_{2}}
\newcommand{\Hs}{\lambda_{3}\lambda_{4};\lambda_{1}\lambda_{2}}
\newcommand{\Jx}{J_{\xi}}
\newcommand{\lamf}{\lam_{\overline{2}}\lam_{4}}
\newcommand{\lami}{\lam_{1}\lam_{\overline{3}}}
\newcommand{\Lamb}{\Lambda_{b}}
\newcommand{\Lamc}{\Lambda_{c}}
\newcommand{\Lam}{\Lambda}
\newcommand{\nn}{\nonumber}
\newcommand{\ra}{\rightarrow}
\newcommand{\pbp}{\overline{p}p}
\newcommand{\ppb}{p\overline{p}}
%___________________________________________________________________________________
%%  No Modify  %%

\fancyhead[co]{\xiaowuhao Cenxi Yuan~ et al: Two-nucleon excitation from $p$ to $sd$  shell in $^{12,14}$C  }    %Write fancyhead in here

\footnotetext[0]{ {\bf Received date:~~~~~~~~~~~~~~~~~}                 \quad {\bf Revised date:}  }

\title{Two-nucleon excitation from $p$ to $sd$  shell in $^{12,14}$C\thanks{{\bf Foundation item:~} The sponsorship with the National Natural Science Foundation
of China under Grant No.~11305272, the Specialized Research Fund for the Doctoral Program
of Higher Education under Grant No.~20130171120014, the Fundamental Research Funds
for the Central Universities under Grant No.~14lgpy29, and the Guangdong Natural Science Foundation under Grant No.~2014A030313217.
} }

\author{%
Cenxi Yuan\xmark{1,\#}, Min Zhang\xmark{1}, Nianwu Lan\xmark{1}, Youjun Fang\xmark{1}
%%%%%%%  Write foonote
\email{{\bf Biography:}\quad Cenxi Yuan(1984-),
male(Chinese), Zhengzhou, Ph.D./Lecturer, Nuclear Physics; yuancx@mail.sysu.edu.cn.  } }

\maketitle

\address{% institution and/or address
(1.~{\it Sino-French Institute of Nuclear Engineering and Technology, Sun Yat-Sen University, Zhuhai \ 519082 \ China}}

\begin{center}

%%%%%%% Write  abstract  in here  %%%%%
\begin{abstract}
The effect of the two-nucleon excitation from $p$ to $sd$ shell is discussed in $^{12}$C and $^{14}$C in the frame work of shell model.
The recently suggested shell-model Hamiltonian YSOX provides an suitable tool to investigate the $2~\hbar\omega$ excitation in $psd$ region.
Because the strength of the $<pp|V|sdsd>$ interaction,  which represents the interaction between the $0~\hbar\omega$ and $2~\hbar\omega$ configurations, is considered in the construction of the YSOX. The level of $^{12}$C is almost independent on the $<pp|V|sdsd>$ interaction, but excitation energies of certain states in $^{14}$C are strongly affected by it. Further investigation shows that the percentage of $2~\hbar\omega$ configuration in these states is quite different from that of the ground state.
\end{abstract}
\end{center}

%%%%%%%% Write keyword in here %%%%%%%%%

%%%%% Please give the information if you are familiar with it. see the introduction to CLC%%
\CLCandDocCode{}{A}   %Fill the Chinese Library Classification index

%%%%%%%%%%%%%%%%%%%%%%%%%
%%%%%%%%%%%%%%%%%%%%%%%%%
\vspace{0.9mm}

\section{Introduction}

The new generation of facilities are used to investigate the proton- and neutron-rich nuclei with the radioactive isotope beam. Such as, the scientists discovered $45$ new neutron-rich nuclei on Radioactive Isotope Beam Factory in 2010~\cite{thoennessen2011}. One of the important issue in nuclear physics is to understand the nuclei from the stability line to the drip line. Thanks to the extensive experimental and theoretical study on the properties of proton- and neutron-rich nuclei in recent years, the position of the drip line is known up to oxygen isotopes~\cite{thoennessen2012}.

In a theoretical view, how to solve the nuclear system is a long standing big challenge. The common theoretical methods are roughly divided into three categories: the \emph{ab initio} methods, the methods based on the mean field approximation, and the nuclear shell model~\cite{brown2001,Caurier2005}. The nuclear shell model solves the many-body Schrodinger equation in a truncated model space. The modern nuclear shell model includes the configuration mixing and the residue interactions. The binding energies and the wave functions of both the ground and excited states can be given simultaneously after the diagonalization process. The observed binding energies, levels, electromagnetic properties, $\beta$ decays, and many other properties can be well describe through shell model approach in light and medium mass region~\cite{brown2001,Caurier2005}.

For light nuclei in $psd$ region, some well determined shell-model Hamiltonians are constructed by fitting to the observed binding energies and levels, such as MK~\cite{mk1975}, WBT and WBP~\cite{wbt1992}. The fitting procedure of these Hamiltonians is limited in the $0-1~\hbar\omega$ model space and considers the strength of $<pp|V|pp>$, $<sdsd|V|sdsd>$ and $<psd|V|psd>$ parts of the interaction. The two-nucleon excitation from $p$ to $sd$ shell and the corresponding $<pp|V|sdsd>$ interaction are not included in the construction of the Hamiltonians. Recently, a new Hamiltonian for $psd$ shell, YSOX, is introduced~\cite{yuan2012}. YSOX well describes the binding energies, levels, electromagnetic properties, and Gamow-Teller transitions of boron, carbon, nitrogen, and oxygen isotopes~\cite{yuan2012}. The construction of the YSOX includes the consideration of the $<pp|V|sdsd>$ interaction, which allows an investigation on the two-nucleon excitation from the $p$ to $sd$ shell.

The percentage of the $2~\hbar\omega$ configuration dramatically decreases from $^{16}$O to $^{24}$O, which indicates its importance in the description of nuclei from the stability line to the drip line~\cite{yuan2012}. It is easy to understand such changes on the isospin degree of freedom. The main configuration $^{16}$O fully occupies the $p$ shell in the independent particle model (IPM). After the inclusion of the residue interaction, the ground state of $^{16}$O shows strong mixing between the $0$ and $2~\hbar\omega$ configurations. Actually, the higher $\hbar\omega$ configuration is also important for the exactly description of $^{16}$O~\cite{brown2001}. From $^{16}$O to $^{24}$O, the valence neutrons in $sd$ shell increase. The mixing between the $0$ and $2~\hbar\omega$ configurations contribute less in the total energies.

The two-nucleon excitation shows its effect on the degree of freedom of excitation energy. In $^{10}$B and $^{17}$C, levels of some excited states are strongly affected by the strength of the $<pp|V|sdsd>$ interaction, but others are almost independent on that interaction. The present work investigate how and why the $2~\hbar\omega$ configuration influence the levels in $^{12}$C and $^{14}$C in the frame work of shell model.

\section{Shell model}
Modern shell model includes the residue two-body interactions and treats the states of nuclei as the configuration mixing of all possible states. In principle, the energies and wave functions can be obtained through solving the Schrodinger equation. But for the majority of nuclei with many protons and neutrons, the model space is too huge. One can limit the cost of the calculations by selecting a core, normally a doubly magic nuclei. The model space is then reduced to several valence proton and neutron orbits. The energies and wave functions can be obtained by solving the Schrodinger equation in the truncated model space.

Because the model space is reduced to the effective model space, a corresponding effective Hamiltonian is needed to be used when solving the Schrodinger equation. Effective Hamiltonian is normally derived from nucleon-nucleon potential in two ways, one is phenomenological by fitting  the nucleon-nucleon potential to the experimental binding energies and energy levels,  the other is realistic by using nucleon-nucleon potential derived from the pion-nucleon scattering and the nucleon-nucleon scattering data.

In this paper, the effective model space is $psd$ space, and corresponding effective Hamiltonians are MK~\cite{mk1975}, WBT~\cite{wbt1992}, WBP~\cite{wbt1992}, and recently suggested YSOX~\cite{yuan2012}. These phenomenological Hamiltonians fit their two-body matrix elements (TBME) to the nuclear structure data, especially the binding energies and energy levels. Shell-model calculations are forwarded through the programme OXBASH~\cite{OXBASH}.

The new Hamiltonian for $psd$ region YSOX~\cite{yuan2012} is developed from $V_{MU}$~\cite{otsuka2010}, SFO~\cite{suzuki2003} and SDPF-M~\cite{utsuno1999}. The $\langle pp|V|pp\rangle$ and $\langle sdsd|V|sdsd\rangle$ parts of TBME are from SFO and SDPF-M, respectively. The $\langle psd|V|psd\rangle$
($\langle pp|V|sdsd\rangle$) TBME are calculated through $V_{MU}$~\cite{otsuka2010} plus M3Y~\cite{m3y1977} spin-orbit force as follows,
\begin{eqnarray}
V = 0.85(0.55)V_{central}+ V_{tensor}(\pi+\rho) + V_{spin-orbit}(\text{M3Y}).
\end{eqnarray}
The $V_{central}+ V_{tensor}(\pi+\rho)$ is the original $V_{MU}$. In the present study, we
reduce the central force in TBME in $\langle psd|V|psd\rangle$ and $\langle
pp|V|sdsd\rangle$ by factors $0.85$ and $0.55$ from
the original $V_{MU}$, respectively. More details can be found in Ref.~\cite{yuan2012}.

\section{Results}

\begin{figure}
   \centering
 \scalebox{0.4}{\includegraphics{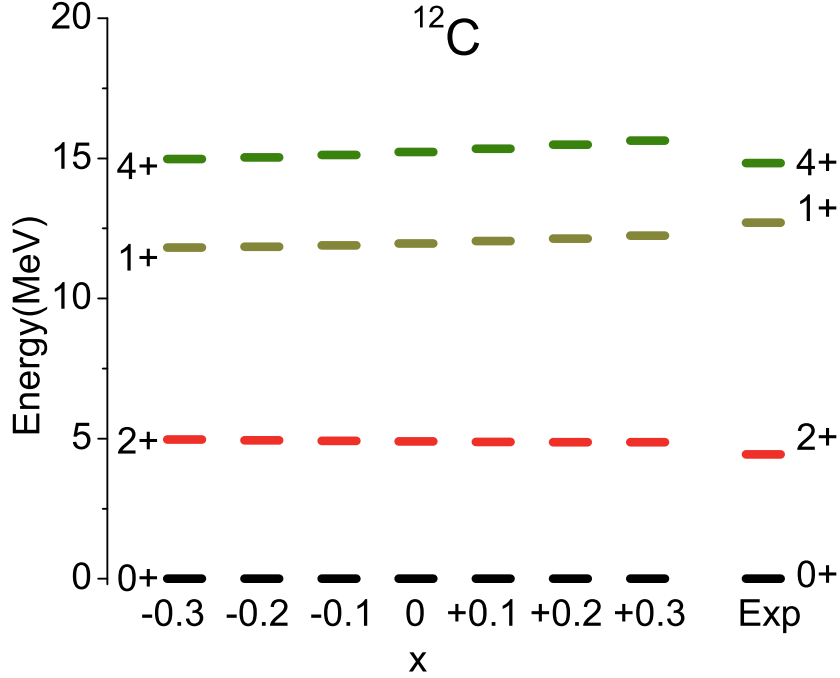}}
   \figcaption{\label{12C}  The levels of $^{12}$C as the function of $x$, which specifies the strength of the $<pp|V|sdsd>$(central)$=(0.55+x)V_{MU}$(central). Experimental data is from NNDC~[13]. }
\end{figure}

\begin{figure}
   \centering
 \scalebox{0.4}{\includegraphics{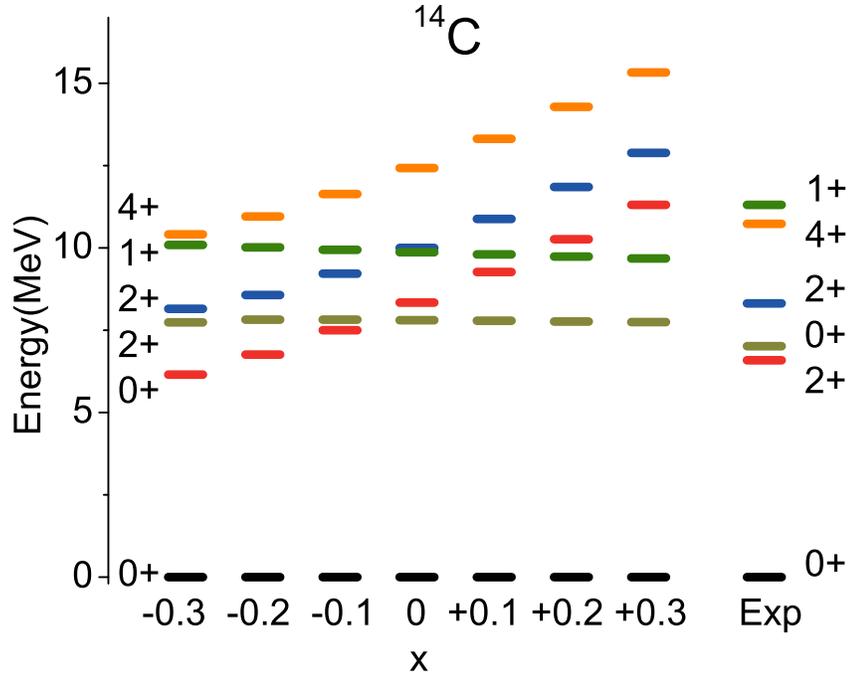}}
   \figcaption{\label{14C}  The same as Fig.~\ref{12C} but for $^{14}$C. }
\end{figure}

\begin{figure}
   \centering
 \scalebox{0.4}{\includegraphics{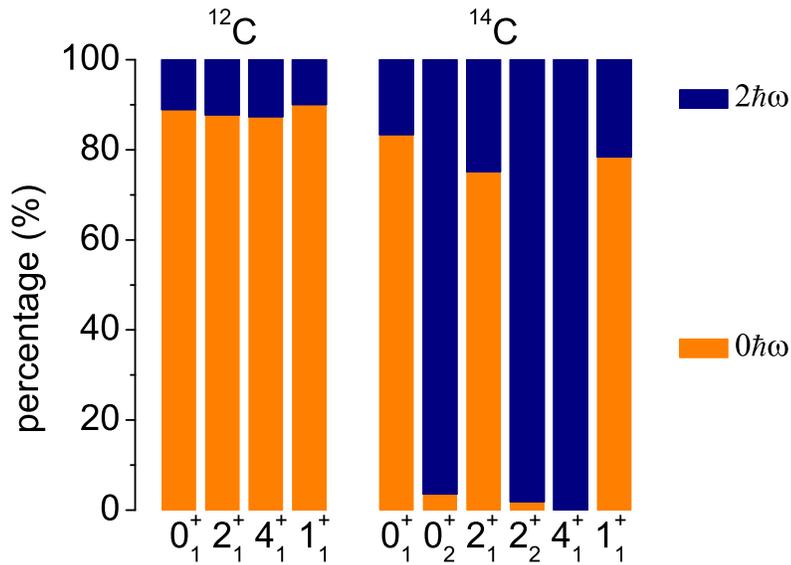}}
   \figcaption{\label{per}  The percentage of the $0~\hbar\omega$ and $2~\hbar\omega$ configuration in ground and excited states of $^{12}$C and $^{14}$C. }
\end{figure}

The percentage of $2~\hbar\omega$ configuration changes in each isotope when neutron number increasing, such as, from $25\%$ in $^{16}$O to $10\%$ in $^{24}$O~\cite{yuan2012}. It may also change from ground to excited states. Ref.~\cite{yuan2012} discussed the effect of the $<pp|V|sdsd>$ interaction on the energy levels of $^{10}$B and $^{17}$C. The $<pp|V|sdsd>$ interaction is the interaction between $0~\hbar\omega$ and $2~\hbar\omega$ configuration. If the excited state has the similar $2~\hbar\omega$ configuration to the ground state, the strength of $<pp|V|sdsd>$ interaction rarely affect the excitation energy, and vice versa. But both in $^{10}$B and $^{17}$C, the excitation energies of some states changes a lot when the strength of $<pp|V|sdsd>$ interaction changes. Such results indicate the percentage of $2~\hbar\omega$ configuration in these excited states are different from those of the ground states. It is interesting to see that some low lying states in $^{10}$B has relatively large $2~\hbar\omega$ configuration. Although the valence nucleons of $^{10}$B are mainly active in the $p$ shell, the $1^{+}_{1}$ state has $16\%$ $2~\hbar\omega$ configuration, around $10\%$ larger than that of the ground state. The previous investigations show that the mixing of $0~\hbar\omega$ and $2~\hbar\omega$ configuration is important in the description of both ground and excited states in both stable and exotic nuclei.

The present work aims to extend the knowledge on the effect of the two-nucleon excitation from $p$ shell to $sd$ shell. Fig.~\ref{12C} and~\ref{14C} present the levels of $^{12}$C and $^{14}$C as the function of the strength of the $<pp|V|sdsd>$ interaction. As discussed before, if the excited states have similar $2~\hbar\omega$ configuration to the ground state, the excitation energies do not change much when the $<pp|V|sdsd>$ interaction changes. In Fig.~\ref{12C}, it is seen that the levels of $^{12}$C changes little as the $<pp|V|sdsd>$ interaction increasing or decreasing. But for $^{14}$C, the excitation energies of $0^{+}_{2}$, $2^{+}_{2}$, and $4^{+}_{1}$ states are dramatically dependent on the $<pp|V|sdsd>$ interaction.

To further investigate the $2~\hbar\omega$ configuration in different states of $^{12}$C and $^{14}$C, the percentage of both $0~\hbar\omega$ and $2~\hbar\omega$ configuration are presented in Fig.~\ref{per}. It is clear seen that the $2~\hbar\omega$ configuration is almost the same in each states of $^{12}$C, but very different in those of $^{14}$C. The valence nucleons of $^{12}$C are mostly excited inside $p$ shell. The two-nucleon excitation from $p$ to $sd$ shell is around $10\%$ in each state and does not change the level of $^{12}$C much. In $^{14}$C, the most important configuration for ground state is $\pi(p_{3/2})^{4}\nu(p_{3/2})^{4}(p_{1/2})^{2}$. The $2^{+}_{1}$ is mainly excited by one protons moving to $p_{1/2}$ orbit, which does not increase or decrease the $2~\hbar\omega$ configuration compared with that of ground state. But for $0^{+}_{2}$ and $2^{+}_{2}$, they are mainly excited by moving two neutrons to $sd$ shell, rather than two protons to $p_{1/2}$ orbit. The $2~\hbar\omega$ configuration becomes the most important configuration ($>95\%$). The $4^{+}_{1}$ is all contributed by $2~\hbar\omega$ configuration. Because the two proton holes in $p$ shell can not couple to the angular momentum $4$.

As a results of the $2~\hbar\omega$ configuration in each state, the levels are affected by the $<pp|V|sdsd>$ interaction. When the strength of the $<pp|V|sdsd>$ interaction is increasing, the ground state, $2^{+}_{1}$, and $1^{+}_{1}$ states are more binding because of the $2~\hbar\omega$ configuration around $20\%$. But for $0^{+}_{2}$, $2^{+}_{2}$, and $4^{+}_{1}$ states, they are (almost) purely $2~\hbar\omega$ configuration. Therefor their binding energies are rarely dependent on the strength of the $<pp|V|sdsd>$ interaction, resulting a changing level relative to ground state.

\section{Summary}
The present work discuss the two-nucleon excitation from $p$ to $sd$ shell in light nuclei based on the framework of shell model. The $2~\hbar\omega$ configuration is important in the investigation of binding energy in a long isotope chain and in the different states in one nucleus. The relationship between the strength of the $<pp|V|sdsd>$ interaction and the levels of $^{12}$C and $^{14}$C are researched through Hamiltonian YSOX. It is shown that the strength of the $<pp|V|sdsd>$ interaction is very influential to the levels of states with different $2~\hbar\omega$ configuration from that of the ground state. The strength of such interaction linked the $0~\hbar\omega$ and $2~\hbar\omega$ configuration is rarely investigated because its effect is not obvious in the levels of many nuclei. It is interesting to further find the nuclei of which the level is sensitive to the strength of the $<pp|V|sdsd>$ interaction, which is helpful for the understanding how the nuclear force driving the nuclear structure.

\section{Acknowledge}
The author acknowledge to the useful suggestions from Furong Xu, Takaharu Otsuka, and Toshio Suzuki.

\clearpage

\end{document}